\documentclass[12pt]{article}
\usepackage{amssymb,amsmath,epsfig,graphicx}
\usepackage{amsfonts}
\usepackage{amsthm}
\usepackage{caption}
\usepackage{epstopdf}

\begin{document}
\title{\bf Gravastars in $f(\mathcal{G},T)$ Gravity}
\author{M. Farasat Shamir \thanks{farasat.shamir@nu.edu.pk}
and Mushtaq Ahmad \thanks{mushtaq.sial@nu.edu.pk}\\\\ National University of Computer and
Emerging Sciences,\\ Lahore Campus, Pakistan.}

\date{}

\maketitle
\begin{abstract}
This work proposes a stellar model under Gauss-Bonnet $f(\mathcal{G}, T)$ gravity with the conjecture theorised by Mazur and Mottola, well known as the gravitational vacuum stars (gravastars). By taking into account the $f(\mathcal{G},T)$ stellar model, the structure of the gravastar with its exclusive division of three different regions namely, (i) the core interior region (ii) the junction region (shell), and (iii) the exterior region, has been investigated with reference to the existence of energy density, pressure, ultra-relativistic plasma, and repulsive forces. The different physical features like, the equation of the state parameter, length of the shell, entropy, energy-thickness relation of the gravastar shell model have been discussed. Also, some other physically valid aspects have been presented with the connection to non-singular and event-horizon free gravastar solutions, which in contrast to a black hole solution, might be stable without containing any information paradox.

\end{abstract}

\section{Introduction}

Gravastars (gravitational vacuum stars) have been proposed by Mazur and Mottola \cite{MZR} featuring the exact solutions of eminent Einstein field equations. The original presentation describes them as super compact, spherically symmetric and singularity free objects, that can be considered to be virtually as compact as the black holes. In the presented three-layer model, this high compactness is reinforced by a de-Sitter spacetime, enclosed by a thin shell made up of some ultra-stiff matter, and the exterior vacuum core is, of course, that of Schwarzschild solution so that the Schwarzschild spacetime singularity is removed. Though, their creation procedure is not yet clear, however, the idea is quite captivating as it could resolve two main fundamental obstacles related to black holes: the singularity problem and the paradoxical problems. Regardless of these theoretical and observational achievements, a number the challenging problems still exist which frequently inspire researchers to look for other substitutes, in which the endpoints of the final phase of gravitational collapse are huge stars without event horizons. Example of such objects, to mention only a few of them, include, black stars \cite{PRD,PRDM}, Bose superfluid \cite{PRD1}, and gravastars that stresses the matter within the bounds of the gravitational radius $r_s = 2GM/c^2$, i.e., much closer to the Schwarzschild radius but with no singularity and event horizon \cite{PRD2}. Among these objects, gravastars have received much attention recently \cite{PRD3}, partly because of the firm connection between the cosmological constant and the so called currently accelerating Universe \cite{PRD4}, although very strict observational limitations on the presence of such stars may occur \cite{PRD5}.

In the originally presented model by Mazur and Mottola \cite{MZR}, the  gravastars comprise of five layers: an inner core $0<r<r_1$ with $\rho=-p$, defined by the de-Sitter spacetime, an intermediary thin layer made of some stiff fluid $r_1<r<r_2$ with $\rho=p$, an exterior region $r>r_2$ with $\rho=p=0$, designated by the Schwarzschild solution, and infinitely two thin shells, apparently, on the regions $r=r_1$ and $r = r_2$, respectively. These characteristics are the most significant one for the construction of a gravastar model with negative central pressure, positive energy density and without event and cosmological horizons.  It may be noted that here $\rho$ is for energy density and $p$ being the isotropic pressure of the gravastar. However, the inner core possesses the constant energy density, given by $\rho_{int}=3H_{0}^{2}/8\pi\geq0$. Physically, the region $r_1<r<r_2$ is the most vital one because this is the region where the non-trivial gravastar model can be specified. Also, it has been suggested that the five-layer models may be simplified to the three-layer ones, by replacing two infinitely thin stiff shells and the intermediary region with just one infinitely thin shell core \cite{Visser3}.

Motivated by the originally proposed gravastar model, numerous works on gravastars have been initiated, but majority of them have only provided the stationary solutions except in \cite{PRD8}. A similar simple model consisting of infinitesimally thin shell configuration was anticipated in \cite{Visser3} and a generalized one in \cite{Carter2}. Some other possibilities for the interior solutions have also been discussed \cite{Bilic}-\cite{LoBO10}. In a recent work \cite{PRD3}, a viable solution for the gravastar having electrically charged configurations has been studied and the bounds on the astronomical data for the existence of the gravastars are proposed \cite{PRD5}.
Kubo and Sakai \cite{PRD6} studied gravitational lensing effects as a possible approach to detect the gravastars  by adopting the Visser-Wiltshire model of spherical thin-shell geometry, which provides the connection of the de-Sitter interior core to the Schwarzschild exterior core over the assumption of optically transparent surface. They calculated the image of the rotating companion around the gravastar and and found some characteristic images, which were dependent on whether the gravastar possessed unstable spherical orbits of photons or not. They determined the total change in luminosity, known as the microlensing effects and concluded that the maximal luminosity could be significantly bigger than the black hole with unchanged mass. For the scenario, the maximum luminosity could be substantially higher than the black hole with the unchanged mass.  Similar astrophysical impacts of such results have been discussed specifically for $f(R,T)$ gravity \cite{PRD7}.
If the gravastars exist in the universe, how can they be recognized just by observations? Chirenti and Rezzolla \cite{PRD13} put a question how to differentiate a gravastar from a black hole. They investigated the axial-perturbations on gravastars and established that the quasi normal modes of gravitational waves were different from those of the black holes. Some more generalized perturbations were studied later by Pani and collaborators \cite{PRD14}.

We have just made a note on the variety of work associated to the gravastars which are mainly based on different physical assumptions followed by mathematical arguments. But, just to mention here, all these works carried out by many authors are mainly confined to the framework of Einstein theory of relativity  \cite{Visser3}-\cite{Rahaman3}. Without any doubt, Einstein theory of relativity has been unparalleled wonderful tool to address the complicated issue related to the accelerated expansion of this universe. But, many of the valid inquiries about the factors responsible for this accelerated expansion of the universe such as dark matter and energy, flatness issue, cosmological constant and initial singularity have remained unanswered by general relativity (GR). Clearly, GR alone fails to tackle these important issues which are understood to be the real reason behind the so called expansion of universe.
The modified theories of gravity have provided some different criteria to solve complexities arisen due to the quantum gravity and have given the researchers with different astronomical strategies to explore the reasons behind this accelerating expansion of universe. Harko and collaborators \cite{HarkoA} were the pioneers who presented the idea of implicit and explicit couplings of  curvature and matter terms in $f(R,T)$ gravity.
Some remarkable works on modified theory of gravity can be noted
in \cite{NJO}-\cite{Sta}.
Recently, researchers have
shown some generalized techniques in modified
Gauss-Bonnet gravity. Sharif and Ikram \cite{sharif.ayesha}
introduced another modified $f(\mathcal{G},T)$ theory of gravity and
determined diverse energy constraints for the Friedmann-Robertson-Walker
(FRW) metric and they concluded that the significant test particles followed
non-geodesic ways of geometry in the presence of some additional dynamical forces.
One may expect some new definitions pertaining to late-time
cosmic acceleration assuming some specific $f(\mathcal{G},T)$
gravity models. We implemented Noether symmetries to find the exact solutions of the modified field equations in $f(\mathcal{G},T)$
theory of gravity \cite{Sir&M}. Also, we explored some cosmological viable $f(\mathcal{G},T)$
gravity models in anisotropic background for locally rotationally symmetric
Bianchi type $I$ universe using the Noehter symmetries \cite{Sir&Me2}. We proposed that some
Gauss-Bonnet gravity specific models may be used for the
reconstruction of $\Lambda$CDM cosmology without indulging any
cosmological constant.
A unique stellar model in $f(R ,T)$ gravity describing some viable features of the gravastars have been worked out by Das et al. \cite{GRVFRT}. By taking into account our previous work on the compact stars in $f(\mathcal{G}, T)$ gravity \cite{Sir&Me3,Sir&SZ}, we are motivated to use $f(\mathcal{G}, T)$ gravity stellar model to investigate the gravastars.

The present setup of this paper is organized as follows: In Section II, the fundamental formalism of $f(\mathcal{G}, T)$ gravity has been given with some important description of characteristics of the modified gravity. Section III, includes the basic spherically symmetric equations with perfect fluid matter distribution. Moreover, it contains the stellar structure expressions presented with the boundary conditions and the equation of state (EoS) parameter for the analysis of the compact stars in $f(\mathcal{G}, T)$ gravity. In Section IV, the geometry of the gravastar, comprised of three exclusive regions, the interior, exterior and the thinly structured shell of the gravastar is discussed with the construction of the gravastar solutions under $f(\mathcal{G},T)$ gravity. Several physical attributes of the stellar model such as EoS parameter, proper length of the thin shell, entropy, and the energy within the shell have been analyzed in section V. Finally, Section VI gives the conclusive discussion on the results with the brief presentation of some important astrophysical implications under $f(\mathcal{G}, T)$ gravity.

\section{$f(\mathcal{G},T)$ Gravity}

 Let us start with an expression for the general action of modified Gauss-Bonnet $f(\mathcal{G},T)$ gravity \cite{sharif.ayesha}
 \begin{equation}\label{action}
\mathcal{U}= \frac{1}{2{\kappa}^{2}}\int d^{4}x
\sqrt{-g}[R+f(\mathcal{G},\mathrm{\textit{T}})]+\int
d^{4}x\sqrt{-g}\mathcal{L}_{\mathcal{M}},
\end{equation}
where $f(\mathcal{G},T)$ is a function comprising of the Gauss-Bonnet term
$\mathcal{G}$ with the trace of the energy-momentum tensor $T$, $g$ gives the metric determinant.
$\kappa$ is for the coupling constant, $R$ is the Ricci Scalar, and $\mathcal{L}_{\mathcal{M}}$
expresses the matter part of the Lagrangian.\par
 The variation of Eq.(\ref{action}) with
respect to the metric tensor $g_{\xi\eta}$, one can have the following field equations
\begin{eqnarray}\nonumber
G_{\xi\eta}&=&\mathrm{\textit{T}}_{\xi\eta}+[2Rg_{\xi\eta}\nabla^{2}+
2R\nabla_{\xi}\nabla_{\eta}+4g_{\xi\eta}R^{\mu\nu}\nabla_{\mu}\nabla_{\nu}+
4R_{\xi\eta}\nabla^{2}-4R^{\mu}_{\xi}\nabla_{\eta}\nabla_{\mu}\\\nonumber
&&-4R^{\mu}_{\eta}\nabla_{\xi}\nabla_{\mu}-4R_{\xi\mu\eta\nu}\nabla^{\mu}\nabla^{\nu}]f_{\mathcal{G}}+
\frac{1}{2}g_{\xi\eta}f-[\mathrm{\textit{T}}_{\xi\eta}+\Theta_{\xi\eta}]f_{\mathrm{\textit{T}}}\\
&&-[2RR_{\xi\eta}-4R^{\mu}_{\xi}R_{\mu\eta}-4R_{\xi\mu\eta\nu}R^{\mu\nu}+2R^{\mu\nu\delta}_{\xi}R_{\eta\mu\nu\delta}]
f_{\mathcal{G}},\label{4_eqn}
\end{eqnarray}
where $\Box=\nabla^{2}=\nabla_{\xi}\nabla^{\xi}$ represents the
d'Alembertian operator, ${G}_{\xi\eta}=R_{\xi\eta}-\frac{1}{2}g_{\xi\eta}R$ gives the
Einstein tensor, $\Theta_{\xi\eta}= g^{\mu\nu}\frac{\delta
\mathrm{\textit{T}}_{\mu\nu}}{\delta g_{\xi\eta}}$, $f\equiv f(\mathcal{G},T)$, $f_{\mathcal{G}}\equiv\frac{\partial f (\mathcal{G},\mathrm{\textit{T}})}{\partial
\mathcal{G}}$, and $f_{\mathrm{\textit{T}}}\equiv\frac{\partial
f(\mathcal{G},\mathrm{\textit{T}})}{\partial \mathrm{\textit{T}}}$.
   Einstein equations can be reawakened by simply substituting
$f(\mathcal{G},\mathrm{\textit{T}})=0$ whereas field equations for
$f(\mathcal{G})$ are reproduced by replacing
$f(\mathcal{G},\mathrm{\textit{T}})$ with $f(\mathcal{G})$ in Eq.($\ref{4_eqn}$).
The matter energy-momentum tensor denoted by $\mathrm{\textit{T}}_{\xi\eta}^{(m)}$ can be defined as
 \begin{equation}\label{emt}
\mathrm{\textit{T}}_{\xi\eta}^{(m)}=-\frac{2}{\sqrt{-g}}\frac{\delta(\sqrt{-g}\mathcal{L}_{M})}{\delta
g^{\xi\eta}}.
\end{equation}
However, the metric dependent matter energy-momentum tensor will have the form
\begin{equation}\label{emt1}
\mathrm{\textit{T}}_{\xi\eta}^{(m)}=g_{\xi\eta}\mathcal{L}_{M}-2\frac{\partial\mathcal{L}_{M}}{\partial
g^{\xi\eta}}.
\end{equation}
The covariant divergence of Eq.(\ref{4_eqn}) is given as
\begin{equation}\label{div}
\nabla^{\xi}T_{\xi\eta}=\frac{f_{\mathrm{\textit{T}}}(\mathcal{G},\mathrm{\textit{T}})}
{\kappa^{2}-f_{\mathrm{\textit{T}}}(\mathcal{G},\mathrm{\textit{T}})}\bigg[(\mathrm{\textit{T}}_{\xi\eta}+\Theta_{\xi\eta})
\nabla^{\xi}(\text{ln}f_{\mathrm{\textit{T}}}(\mathcal{G},\mathrm{\textit{T}}))+
\nabla^{\xi}\Theta_{\xi\eta}-
\frac{g_{\xi\eta}}{2}\nabla^{\xi}T\bigg].
\end{equation}
There is a possibility that this theory of gravity might be overwhelmed by the effects of the divergences.  These divergences are caused due to the presence of the stress-energy tensor higher order derivatives which are naturally involved in the field equations. This has emerged as an issue to all such modified theories of gravity which include these higher order stress-energy tensor derivatives. The weak equivalence principle is not compromised by the alterations of Einstein theory through the addition of auxiliary field. Though, one may define some new limitations to this equation to attain standard conservation equation for the stress-energy tensor.

The following is the energy-momentum tensor $T_{\xi\eta}$ which defines the perfect fluid resource as
\begin{equation}\label{12}
T_{\zeta\eta}=(\rho+p)\xi_{\zeta}\xi_{\eta}-{p}g_{\zeta\eta},
\end{equation}
where $p$ and $\rho$ are the pressure and energy density, respectively and $\xi_{\zeta}$ denotes
the 4-velocity vector. It is to mention here that we take the lagrangian matter $\mathcal{L}_{(M)}=-p$ which when used in the expression for $\Theta_{\zeta\eta}$, gives
\begin{equation}\label{Bigtheta}
\Theta_{\zeta\eta}=-2T_{\zeta\eta}-p g_{\zeta\eta}.
\end{equation}
Specifically for this study, we choose here the  $f(\mathcal{G},T)$ model as
\begin{equation}\label{VM}
f(\mathcal{G},T)=f_{1}(\mathcal{G})+f_{2}(T),
\end{equation}
where $f_{1}(\mathcal{G})$ is a function comprised of $\mathcal{G}$, such that $f_{1}(\mathcal{G})=\alpha\mathcal{G}^n$, a power law model of $f(\mathcal{G})$ gravity suggested by Cognola et al. \cite{17}, where $\alpha$ is an arbitrary real constant, and $n$ a positive real number.  Here for the second constituent of the model, we have taken $f_{2}({T})=\lambda\textit{T}$, with $\lambda$ a real constant. To keep things simple, we set $\alpha=n=1$ for our upcoming investigations. Now making use of this model in field equation (\ref{4_eqn}), we get
\begin{equation}\label{modelEQ}
G_{\zeta\eta}=(\lambda+1)T_{\zeta\eta}+\frac{1}{2}g_{\zeta\eta}[\lambda(2p+T)+\mathcal{G}].
\end{equation}
It is worth mentioning here that one can retrieve the general field equation in GR by simply putting $\alpha=\lambda=0$. Now for the same  $f(\mathcal{G},T)$ gravity model, Eq.(\ref{div}) takes the new form as
\begin{equation}\label{modiv}
\nabla^{\zeta}T_{\zeta\eta}=\frac{\lambda}{\lambda+1}\bigg[\nabla^{\zeta}(pg_{\zeta\eta})+\frac{1}{2}g_{\zeta\eta}\nabla^{\zeta}T\bigg].
\end{equation}
It is easy to see that by substituting $\lambda=0$ in Eq.(\ref{modiv}), one can get the original conserved form of the energy-momentum tensor in GR.

\section{Modified Field Equations with Spherical Symmetry}

The super-dense cold stars rapidly undergo gravitational collapse due to their overwhelming mass exceeding some critical value. It is impossible to avoid or halt this collapse for the matter with the existing high density EoS parameter. This situation makes one agree on a consensus that a collapsing object must reach within the proper finite time at some singular condition, which is conceptualized as a black hole. Event horizon is the distinguished character which is owned by a black hole. For the sake of simplicity, we consider here an uncharged and non-rotating static, spherically symmetric space-time describing the Schwarzschild black hole as
\begin{equation}\label{11}
ds^{2}=e^{\psi(r)}dt^{2}-e^{-\varphi(r)}dr^2-r^{2}d\theta^{2}-r^2sin^{2}\theta d\phi^{2},
\end{equation}
where $\psi(r)$ and $\varphi(r)$ are the arbitrary radial functions. For this specific situation, these are related as
\begin{equation}\label{horison}
\psi(r)=\frac{1}{\varphi(r)}=1-\frac{r_{s}}{r},~~~~r_{s}\equiv\frac{2GM}{c^2}.
\end{equation}
It can be seen that the  metric $(\ref{11})$ turns out to be singular at $r=r_{s}$, and at $r=0$. Naturally, these two singularities are not the same. In fact, the one appearing at $r=0$ is of generic nature where the spacetime curvature diverges, while the other at $r = r_{s}$ is a coordinate one and may be worked out to disappear after the coordinate is properly transformed. Now if $M$ is sufficiently large and $r=r_{s}$, the local curvature term exhibits the regular behavior, classical test point particle will be freely falling through $r = r_{s}$ and nothing catastrophic happens to it. This particular surface is generally known as an event horizon. The investigations of the properties of such horizons are primarily important, and the studies point out that there may be some deep connection between the gravity and thermodynamics \cite{PRD9}. The finding of the quantum Hawking radiation \cite{PRD10} and the entropy of black hole being proportional to the event horizon area of the black hole further encourages this idea \cite{PRD11}. This fascinating relation was first demonstrated at the time when Jacobson developed the Einstein field equations using the first law of thermodynamics by assuming the proportional relation between the entropy and the area of all local acceleration horizons \cite{PRD12}.

Under these conditions, now it seems favorable to determine regular coordinates extending analytically Schwarzschild exterior geometry through this event horizon to the interior region.
For the metric (\ref{11}), the non-zero components of $G_{\zeta\eta}$ can be found as
\begin{equation}\label{G00}
  G_{00}=\frac{e^{\psi -\varphi } \left(r \varphi '+e^{\varphi }-1\right)}{r^2},
\end{equation}
\begin{equation}\label{G11}
G_{11}= \frac{r \psi '-e^{\varphi }+1}{r^2},
\end{equation}
\begin{equation}\label{G22}
G_{22}=\frac{1}{4} r e^{-\varphi } \left(-r \varphi ' \psi '+2 r \psi ''+r \left(\psi '\right)^2-2 \varphi '+2 \psi '\right).
\end{equation}
Here the prime appearing in the above equations gives derivative with respect to the radius $r$. Now using Eqs.(\ref{12}), (\ref{modelEQ}) and (\ref{G00}-\ref{G22}), one can have the following expressions
\begin{eqnarray}\nonumber
&&2+r^{2}(p\lambda-(2+3\lambda)\rho)=\frac{e^{-2\varphi}}{4}[r^{2}\psi'^{4}-2r^{2}\psi'^{3}\varphi'+8e^{\varphi}(-1+r\varphi'+4\psi'')+\\\nonumber
&&4\psi''(-8+r^{2}\psi'')-4\psi'\varphi'(4(-3+e^{\varphi})+r^{2}\psi'')+
\psi'^{2}(-8+16e^{\varphi}+r^{2}\varphi'^{2}+4\psi'')]\\\label{G0}\\\nonumber,
&&2+e^{\varphi}(-2+r^{2}(-p(2+3\lambda)+\lambda\varphi))=\frac{-e^{-\varphi}}{4}[-r^{2}\psi'^{4}+2r^{2}\psi'^{3}\varphi'-8\varphi'^{2}-\\\nonumber
&&4\varphi''(-8+8e^{\varphi}+r^{2}\psi'')-\psi'^{2}(16(-1+e^\varphi)+r^{2}(\varphi'^{2}+4\psi''))+4\psi'(2e^{\varphi}r+\\\label{G1}
&&\varphi'(4(-3+e^{\phi})+r^{2}\psi''))],\\\nonumber
&&8+e^{\varphi}(-4+r^{4}(-p(2+3\lambda)+\lambda\rho))=\frac{-e^{-\varphi}}{2}[8+r^{2}(6\psi'^{2}-24\psi'\varphi'-\\\label{G2}
&&2\varphi'^{2}+16\psi''+e^\varphi ((2r+(-8+r^{2})\psi')(\psi'-\varphi')+2(-8+r^{2})\psi''))].
\end{eqnarray}
Now for the $f(\mathcal{G},T)$ gravity, the non-conservation Eq.(\ref{div}) of energy-momentum tensor takes the new shape as
\begin{eqnarray}\label{Ndiv}
0=\frac{dp}{dr}+\frac{\psi'}{2}(\rho+p)+\frac{\lambda}{\lambda+1}(p'-\rho').
\end{eqnarray}
In fact, in GR, the Schwarzschild approach has remained pioneer in helping us to select from different possibilities of the matching conditions  while exploring  stellar compact stars. Now for our concerns in case of alternative theories of gravity, the modified Tolman-Oppenheimer-Volkoff (TOV) equations with null pressure and energy density, the exterior solution of the star may differ from the Schwarzschild's solution \cite{OV}. Though,
it is anticipated that the modified TOV solutions  with pressure and energy density  (may be non-zero) may admit Schwarzschild's solutions with some special selection of $f(\mathcal{G},T)$ gravity model. Perhaps, this is the reason that Birkhoff's theorem may not be fulfilled by modified gravity. The detailed investigation of this phenomenon and to comprehend the related issues in context of modified $f(\mathcal{G},T)$ gravity might be an interesting pursuit. Many researchers have taken into account the Schwarzschild solutions for this issue and concluded some interesting results \cite{27a}-\cite{Ast}.
Therefore, it seems quite interesting here to develop the TOV equation for $f(\mathcal{G},T)$ gravity. For this, we consider the gravitational mass $m(r)$ of some sphere with interior radius $r$ such that $e^{-\varphi}=1-\frac{2m}{r}$. Now manipulating Eqs.(\ref{G0}) and (\ref{G1}) together with the preceding gravitational mass-radius relation, one can have
\begin{equation}\label{AD}
  \psi'=-\frac{\frac{2 m^2}{r-2 m}+r m'}{r^4}-\frac{(\lambda +1) r^2 (p+\rho )}{2 m-r},
\end{equation}
Using Eqs.(\ref{G0}-\ref{G2}) along with $e^{-\varphi}=1-\frac{2m}{r}$, the radial derivative of the gravitational mass within the sphere of radius $r$ reads
\begin{equation}\label{Gmass}
 \frac{dm}{dr}= \frac{2 m \left(m+r^3\right)-(\lambda +1) r^6 (p+\rho )}{r \left(2 m+2 r^3-r\right)}.
\end{equation}
Solving together Eqs.(\ref{AD}) and (\ref{Ndiv}), one gets an expression for the hydrostatic equilibrium in $f(\mathcal{G},T)$ gravity as
\begin{eqnarray}\label{mtov}
\frac{dp}{dr}=\frac{-1}{2}(\rho+p)\frac{\frac{m}{r^{4}}-r^{2}\frac{(1+\lambda)(p+\rho)}{(2m-r)}-\frac{\frac{dm}{dr}}{r^{3}}}{(1-\frac{d\rho}{dp})(1+\frac{\lambda}{1-\lambda})}.
\end{eqnarray}
We have imposed here the condition that the energy density $\rho$ is dependent on the pressure $p$ such that $\rho=\rho(p)$. It is noted here that by taking $\lambda=0$, one can work out the TOV equation related to the case of GR.\par
It may become a tempting pursuit when one goes for the simultaneous solutions of the differential Eqs.(\ref{Gmass}) and (\ref{mtov}) for the investigations of some important physical features of the compact stars, though, the detailed  analysis may not be possible for the moment. The integration of these equations requires some specific boundary conditions starting with values $r=0$, in the center of the star \cite{PRD21}:
\begin{equation}\label{BCs}
  m(0)=0,~~~\rho(0)=\rho_{c},~~~p(0)=p_{c}.
\end{equation}
The star surface ($R=r$) is evaluated when $p(R)=0$ where the solution of the interior of the star is smoothly matched to the Schwarzschild exterior solution. The connection between the potential metrics of interior-exterior spacetime is defined by $e^{\psi(R)}=e^{-\varphi(R)}=1-\frac{2M}{R}$, with $M$ being the total stellar mass of the star. Now for the purpose to make these system of equations closed, the role of the EoS parameter becomes significant here. The coupled equations (\ref{Gmass}) and (\ref{mtov}) are now set to be solved numerically (Runge-Kutta $4$th order method may be more appripriate) for the three unknown radial functions $m$, $\rho$, and $p$, after we have defined the EoS. Like the analysis of compact stars in $f(R,T)$ theory of gravity, the investigations of the equilibrium configurations of the stars in $f(\mathcal{G}, T)$ gravity may be initiated by considering the polytropic and MIT bag EoS based models. For the simplest choice, the work by Tooper \cite{PRD22} may be followed by considering $p=\omega\rho^{5/3}$, with $\omega$ being the EoS parametric constant or in particular, one may choose $\omega=1.4745\times10^{-3}[fm^3/MeV]^{2/3}$  \cite{PRD23,PRD24}.

Our next section is dedicated to the construction of solutions of gravastars in $f(\mathcal{G},T)$ gravity in connection to its different regions that is (i) the exterior region (ii) the interior region, and (iii)the shell.
\begin{figure}[hp]
\centering
\captionsetup{justification=centering,margin=2cm}
\begin{tabular}{cccc}
\\ &
\epsfig{file=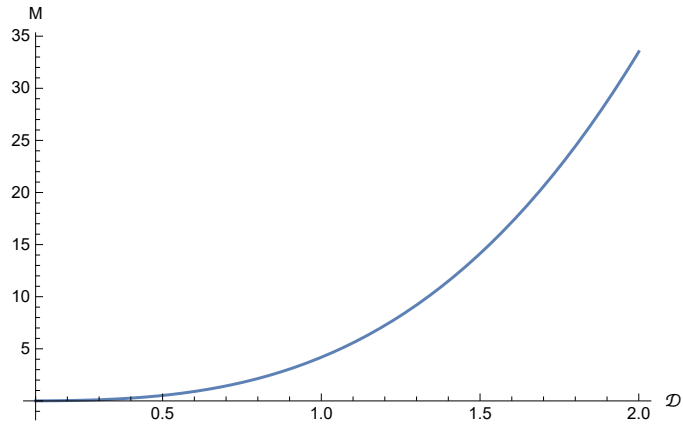,width=0.65\linewidth}\\
\end{tabular}
\caption{ The plot of the mass $M$ $(km)$ with respect to the radius $\mathcal{D}$ $(km)$ of the gravastar with $d_{o}=1$\label{fig:MR}}\center
\end{figure}

\section{Geometry of Gravastars}

We are interested here for the geometrical interpretation and corresponding solutions for the gravastar under investigation. It is not difficult to visualize that the interior of the star is bordered by a very thin shell believed to be made of some ultra-relativistic matter whereas the outer region is entirely vacuum. Therefore, the Schwarzschild's metric can be assumed feasible for this region. As for the geometry of the shell is concerned, it is supposed to be extremely thin having a finite width within the range of $r_{1}=\mathcal{D}\leq{r}<\mathcal{D}+\varepsilon=r_{2}$, where $r_{1}$ and $r_{2}$ describe interior and the exterior radii, respectively of the gravastar under investigation. These regions, in fact, are structured on the basis of EoS parameter as follows: Interior region $\mathcal{R}_1\equiv{0\leq{r}<r_{1}}$, with $\rho=-p$, the shell region $\mathcal{R}_2\equiv{r_{1}}\leq{r}\leq{r_{2}}$, with $+\rho=p$, and the exerior region $\mathcal{R}_3\equiv{r_{2}<r}$, with $\rho=p=0$.

\subsection{The Interior Region}

As discussed above, let us take here EoS parameter as $\rho=\omega{p}$. Now since we have $\omega=-1$ for the interior region $\mathcal{R}_1$, the condition on the EoS parameter gives $p=-\rho$ which describes EoS for the dark energy. Manipulating Eqs.$(\ref{G0})$ and $(\ref{G1})$, one can obtain
\begin{eqnarray}\label{IE}
\rho+p=\frac{e^{-2\varphi}(e^{\varphi}
r+\psi'-\varphi')(\psi'+\varphi')}{r^{2}(1+\lambda)}.
\end{eqnarray}
Imposing the condition of $\rho=-p$ on Eq.(\ref{IE}), and then solving, gives
\begin{eqnarray}\label{IES}
e^{-\varphi}=\frac{C_{1}-r^{2}}{4}~~~\text{and}~~~e^{\psi}&=&C_{2}e^{-\varphi},
\end{eqnarray}
here $C_{1}$ and $C_{2}$ are the constants of integration. A proportionate relationship between the metric potentials $\psi$ and $\varphi$ can be noted from Eq.(\ref{IES}.) It can be seen as well from these solutions that the space-time is independent of any central singularity.
Now extending the same condition of $\rho=-p$ to Eq.$(\ref{Ndiv})$, gives
\begin{equation}\label{CDP}
  \frac{dp}{dr}\big(\frac{2\lambda}{\lambda+1}\big)=0.
\end{equation}
The solution of the above ordinary differential equation (ODE) implies that for $\rho=-p$, we have $\rho=d_{o}$, and $p=-d_{o}$, where $d_{o}$ is a constant of integration. Moreover, one can find the interior gravitational mass ${M}({\mathcal{D}})$ as given below
\begin{equation}\label{M(D)}
{M}({\mathcal{D}})=\int_{0}^{\mathcal{D}}4\pi{r^2}d_{o} dr=\frac{4}{3}\pi{d_{o}}{\mathcal{D}^{3}}.
\end{equation}
It can be noted that with the constant density, there is a directly proportionate mass-radius relationship as shown in Fig.\ref{fig:MR}. This characteristic feature, in fact, is owned by the stellar compact objects.

\subsection{The Exterior Region}

We meet here with another vacuum region $\mathcal{R}_3$ case when the exterior region approaches the flat spacetime as $r$ goes beyond bounds. Obeying the EoS $0=\rho=p$, this exterior region admits the only known Schwarzschild exterior solution
\begin{equation}\label{schwz}
ds^{2}=\Big(1-\frac{2M}{r}\Big)dt^{2}-\Big(1-\frac{2M}{r}\Big)^{-1}dr^{2}-r^{2}(d\theta^{2}+sin^{2}\theta
d\phi^{2}),
\end{equation}
here $M$ is the constant of integration and stands for the total gravitational mass, such that
\begin{equation}\label{schwzrz}
\psi(r)=\frac{1}{\varphi(r)}=1-\frac{2M}{r},~~~~~~~~~r_2\leq{r}.
\end{equation}

\subsection{The Shell Region}

\begin{figure}[hp]
\centering
\captionsetup{justification=centering,margin=2cm}
\begin{tabular}{cccc}
\\ &
\epsfig{file=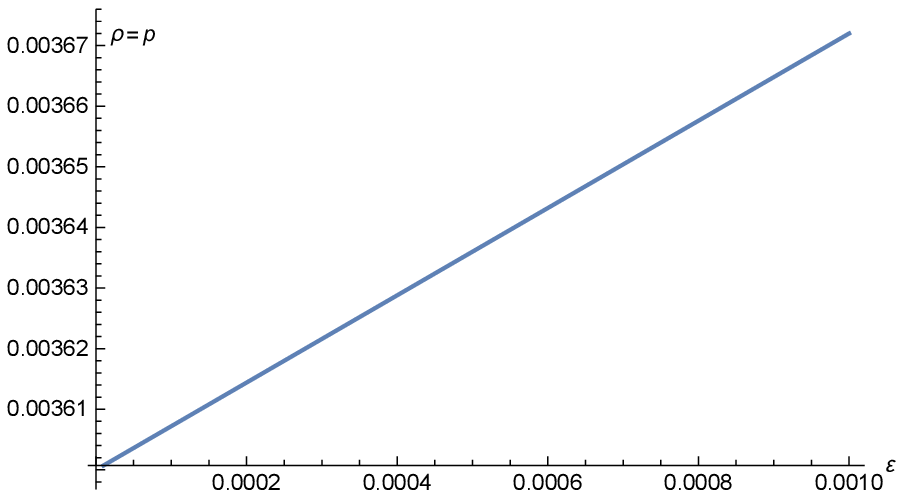,width=0.65\linewidth}\\
\end{tabular}
\caption{ The plot of the density-pressure, $\rho=p$  ($km^{-2}$) of ultra-relativistic stiff fluid in the gravastar shell against the thickness $\varepsilon$ ($km$) of the shell.\label{fig:RPr}}\center
\end{figure}

The only non-vacuum region is $\mathcal{R}_2$ which describes the shell of the gravastar admitting the condition $\rho=p$. It is considered to be consisting of some ultra-relativistic matter connecting to the stiff fluid of cold baryonic universe. In our case, it may originate from the thermal stimulations with no chemical potential or at negligible temperature, it may emerge from the gravitational quanta. If we just see the field Eqs.(\ref{G0}-\ref{G2}), one can conceptualize  that with in non-vacuum shell region $\mathcal{R}_2$, it becomes difficult to find their solutions with the constraint $\rho=p$. One thing which physically establishes a framework for the existence of analytic solutions, is the limit of the thin shell, that is, $0<e^{-\varphi(r)}\ll{1}$. This in fact must be because of the intermediate thin region (thin shell) formed due to the joining of the Schwarzschild exterior vacuum spacetime with that of the vacuum interior spacetime. This thin shell structure suggests that as $r$ approaches to zero, the corresponding radial parameters generally become $\ll1$. Under these newly developed circumstances, and with the implication of the EoS, $\rho=p$, one can manipulate the Eqs.(\ref{G0}-\ref{G2}) to obtain
\begin{equation}\label{DE1}
C_{3}r^{2}=e^{-\varphi}(r^{2}-2\psi')~~~~~\text{and}~~~~~ C_{4}e^{\varphi}=(r-4\psi').
\end{equation}
On solving these two ODEs, one can have
\begin{eqnarray}\label{}
{\psi}={\frac{1}{16}\bigg(2r^{2}+\frac{2\sqrt{2}ArcTanh(\frac{\sqrt{2}r\sqrt{C_{6}}}{\sqrt{C_{5}}})C_{5}^{\frac{3}{2}}}{C_{6}^{\frac{3}{2}}}
+\frac{(-4r+\log(C_{5}-2r^{2}C_{6}))C_{5}}{C_{6}}\bigg)},
\end{eqnarray}
\begin{eqnarray}\label{S2}
e^{-\varphi}=\frac{C_{5}-2r^{2}C_{6}}{8r-16r^{2}}.
\end{eqnarray}
Again implementing the same condition of $\rho=p$ together on Eqs.$(\ref{Ndiv})$ and (\ref{S2}), results into another linear ODE of the form
\begin{equation}\label{ODE2}
\frac{p r^2 \left(C_6 r-C_5\right)}{2 \left(4 C_6 r^2-2 C_5\right)}+p'=0,
\end{equation}
and the solution of which yields the expressions for $\rho$ and $p$ as
\begin{eqnarray}\label{Rhp}
p=\rho=C_{7}e^{\frac{1}{32}\big(-2r^{2}-\frac{2\sqrt{2}ArcTanh(\frac{\sqrt{2}r\sqrt{C_{6}}}{\sqrt{C_{5}}})C_{5}^{\frac{3}{2}}}{C_{6}^{\frac{3}{2}}}
+\frac{(4r-\log(C_{5}-2r^{2}C_{6}))C_{5}}{C_{6}}\big)}.
\end{eqnarray}
Here all the $C_{i}$'s appearing in the above equations are the constants of integration and the radius $r$ corresponds to the shell structure in region $\mathcal{R}_2$ which ranges from $\mathcal{D}+\varepsilon\geq{r}\geq\mathcal{D}$. Therefore, for this particular situation, we have $\varepsilon\ll1$ due to the assumption that $e^{-\varphi}\ll1$.
By further studying Eq.(\ref{Rhp}) with the help of the Fig.\ref{fig:RPr}, it is noted that $\rho\propto{\varepsilon}$ which helps us to conclude that the ultra-relativistic fluid is denser at the outer boundary as compared to the inner boundary of the shell.

\subsection{The Interior-Exterior Joint Interface}

The division of the different regions of the gravastar has already been described by $\mathcal{R}_1$ (the interior region), $\mathcal{R}_{2}$ (the shell region, a joint interface of the interior and the exterior regions), and $\mathcal{R}_{3}$ (the exterior region). A very important formalism used by Darmois \cite{Darmois4} and Israel \cite{Israel2} suggests a smooth match between the interior and exterior regions. At $r=\mathcal{D}$ (the joint surface, $\Xi$), the coefficients of the metric
\begin{equation}\label{NMet}
  ds^{2}=\mathcal{F}(r)dt^{2}-\frac{dr^2}{\mathcal{F}(r)}-r^2(d\theta^{2}+sin^{2}\theta{d\phi^{2}}),
\end{equation}
 are continuous, though the existence of their derivative at this surface may not be guaranteed. However, with the help of Darmois-Israel formalism, it is quite possible to find the expression for the stress-energy surface $\mathcal{S}^{\mu}_{\nu}$. Now for the intrinsic stress-energy surface tensor $\mathcal{S}_{\mu\nu}$, we have by Lanczos equation \cite{Lanczos}
\begin{equation}\label{LCS}
 \mathcal{S}^{\mu}_{\nu}=-\frac{1}{8\pi}(\eta^{\mu}_{\nu}-\delta^{\mu}_{\nu}\eta^{i}_{i}),
\end{equation}
where the tensor $\eta_{\mu\nu}=\Upsilon^{+}_{\mu\nu}-\Upsilon^{-}_{\mu\nu}$ gives the discontinuous surfaces in the extrinsic curvatures or in the surfaces where the second fundamental forms are applicable. It is to be noted that the appearance of $+$ and $-$ symbols describe, respectively, the interior and exterior surfaces. Having two-sided association with the shell, one can express the second fundamental forms as \cite{67}
\begin{equation}\label{LCS1}
 \Upsilon^{\pm}_{\mu\nu}=-\zeta^{\pm}_{i}\bigg[\frac{\partial^{2}x_{i}}{\partial{\xi^{\mu}}\xi^{\nu}}+\Gamma^{i}_{\alpha\beta}\frac{\partial{x^{\alpha}}}{\partial{\xi}^{\mu}}\frac{\partial{x^{\beta}}}{\partial{\xi}^{\nu}}\bigg]|\Xi,
\end{equation}
with $\xi^{\mu}$ being the shell intrinsic coordinates and $\zeta^{\pm}_{i}$ represents the double-sided unit normals on the surface $\Xi$ to the metric (\ref{NMet}), and is expressed as
\begin{equation}\label{nv}
  \zeta^{\pm}_{i}=\pm\bigg|g^{\alpha\beta}\frac{\partial{\mathcal{F}}}{\partial{x^{\alpha}}}\frac{\partial{\mathcal{F}}}{\partial{x^{\beta}}}\bigg|^{-\frac{1}{2}}\frac{\partial{\mathcal{F}}}{\partial{x^i}},~~~~\text{with}~~\zeta^{j}\zeta_{j}=1.
\end{equation}
Now the stress-energy surface tensor can be obtained as $\mathcal{S}_{ij}=diag(\varrho, -\sigma, -\sigma, -\sigma)$ by using the Lanczos equations, with $\varrho$ being the surface energy density and $\sigma$ the surface pressure. Following are the respective expressions for both $\varrho$ and $\sigma$ as
\begin{equation}\label{SED}
\varrho=-\frac{1}{4\pi\mathcal{D}}\big[\sqrt{\mathcal{F}}\big]^{+}_{-},~~~~~
\sigma=-\frac{\varrho}{2}+\frac{1}{16\pi}\bigg[\frac{\mathcal{F^{'}}}{\sqrt{\mathcal{F}}}\bigg]^{+}_{-}.
\end{equation}
Therefore, by using Eq.(\ref{SED}), the new expressions for ϱ and σ read as
\begin{equation}\label{SEDE}
\varrho=\frac{1}{4\pi{\mathcal{D}}}\bigg[\sqrt{\frac{C_{1}-\mathcal{D}^{2}}{4}}-\sqrt{1-\frac{2M}{\mathcal{D}}}\bigg],
\end{equation}
\begin{equation}\label{Sigma}
\sigma=\frac{1}{8\pi}\bigg[\frac{(1-\frac{M}{\mathcal{D}})}{\sqrt{1-\frac{2M}{\mathcal{D}}}}-\frac{C_{1}-2\mathcal{D}^{2}}{2\sqrt{C_{1}-\mathcal{D}^{2}}}\bigg].
\end{equation}
Now using $\varrho$ from Eq.(\ref{SEDE}), we can also express the thin shell mass as
\begin{equation}\label{shm}
 m_{shell}=4\pi\varrho\mathcal{D}^{2}=\mathcal{D}\bigg[\sqrt{\frac{C_{1}-\mathcal{D}^{2}}{4}}-\sqrt{1-\frac{2M}{\mathcal{D}}}\bigg].
\end{equation}
If mass of the shell is known, one can find $M$, the total mass of the gravitational vacuum star by rearranging  Eq.(\ref{shm}) as
\begin{equation}\label{TM}
  M=\frac{1}{8}(4-C_{1})\mathcal{D}+\frac{\mathcal{D}^3}{8}+\frac{1}{2}m\sqrt{C_{1}-\mathcal{D}^2}-\frac{m^{2}}{2\mathcal{D}},
\end{equation}
where $m=m_{shell}$, the mass of the shell of the gravastar.

\section{Some Physical Aspects of the Model}

Since we have discussed in the preceding section the implications of the Lanczos equations and have found the expressions for the surface energy density $\varrho$ and the surface pressure $\sigma$, now we discuss here the important features like the EoS parameter, length of the shell, Entropy, and the energy-thickness relation which exclusively describe the geometry of the gravastar shell.
\subsection{The EoS Parameter}
At the region $r=\mathcal{D}$, the EoS parameter $\omega(\mathcal{D})$ can be written as
\begin{equation}\label{EoS}
  \omega(\mathcal{D})=\frac{\sigma}{\varrho}.
\end{equation}
Now making use of Eqs.($\ref{SEDE}$) and ($\ref{Sigma}$), gives an explicit expression for the EoS parameter as
\begin{equation}\label{EoSp}
  \omega(\mathcal{D})=\frac{\mathcal{D}\bigg[\frac{(1-\frac{M}{\mathcal{D}})}{\sqrt{1-\frac{2M}{\mathcal{D}}}}-\frac{C_{1}-2\mathcal{D}^{2}}{2\sqrt{C_{1}-\mathcal{D}^{2}}}\bigg]}{2\bigg[\sqrt{\frac{C_{1}-\mathcal{D}^{2}}{4}}-\sqrt{1-\frac{2M}{\mathcal{D}}}\bigg]}.
\end{equation}
The above equation comprises of different factors involving the fractions with square root terms which increase the sensitivity of the equation. So, one needs to introduce some restrictions for these terms to keep $\omega(\mathcal{D})$ real. Therefore, with the implications of the conditions $\frac{M}{\mathcal{D}}\ll1$ and $C_{1}-\mathcal{D}^{2}>0$, the expansion of Eq.($\ref{EoSp}$) up to first order with binomial series may give us the positive and the negative values of the EoS parameter.

\subsection{Proper Shell Length}

We have already assumed that the geometry of the shell of the gravastar is at the surface $\mathcal{D}=r$ which is explicitly described by the region $\mathcal{R}_{1}$. Since proper length of the shell is taken as very small such that $\varepsilon\ll1$, thus $r=\mathcal{D}+\varepsilon$ is the interface from where the region $\mathcal{R}_{3}$ starts. This conceptualizes the thickness of these two interfaces. Therefore, the proper length of the shell denoted by $\mathcal{L}$ is calculated as \cite{GRVFRT}
\begin{equation}\label{Lngth}
  \mathcal{L}=\int_{\mathcal{D}}^{\mathcal{D}+\varepsilon}\sqrt{e^{\varphi}}dr=\int_{\mathcal{D}}^{\mathcal{D}+\varepsilon}\frac{2 \sqrt{2}}{\sqrt{\frac{C_6 \left(C_5-2 C_6 r^2\right)}{C_5 (1-2 r) r}}}dr.
\end{equation}
Integration of Eq.(\ref{Lngth}) gives us
\begin{eqnarray}\nonumber
\mathcal{L}&=&\Bigg[\frac{1}{\sqrt{2}\sqrt{1+\frac{1}{1-2r}+\frac{1}{r}}r(-1+2r)}(1+\sqrt{2})\Big\{4r(-1+\sqrt{2})(-1+2r^{2})\\\nonumber
&+&2^{\frac{1}{4}}\sqrt{-(1-2r)^{2}}\sqrt{-\sqrt{2}+2r}
\sqrt{\frac{\sqrt{2}+2r}{1-2r}}\Big\{2^{\frac{1}{4}}\sqrt{(-10+7\sqrt{2})r}EllipticE\\\nonumber
&\times&\Big[\arcsin\Big[\frac{\sqrt{-2+\sqrt{2}}r}{\sqrt{1-2r}}\Big],3+2\sqrt{2}+\sqrt{r-\sqrt{2}
r}\Big((-1+\sqrt{2})EllipticF\\\nonumber
&\times&[\arcsin\Big[\frac{2^{\frac{1}{4}}\sqrt{r-\sqrt{2}
r}}{\sqrt{1-2r}}],3+2\sqrt{2}\Big]+Elliptic\pi[2+\sqrt{2},\\\label{LoS}
&\times&\arcsin\Big[\frac{2^{\frac{1}{4}}\sqrt{r-\sqrt{2}
r}}{\sqrt{1-2r}}\Big],3+2\sqrt{2}]\Big)\Big]\Big\}\Big\}\Bigg]^{\mathcal{D+\varepsilon}}_{\mathcal{D}},
\end{eqnarray}
where the special functions $EllipticE$, $EllipticF$, and $Elliptic\pi$ appearing in the above equation give the complete elliptic integral $E(m)$, elliptic integral of the second kind $E(\phi)/m$, and complete elliptic integral of the third kind $\Pi(n/m)$ respectively, with $m$ and $n$ being the parameters. Any elliptic integral can be expressed in terms of the three standard kinds of Legendre-Jacobi elliptic integrals.
\begin{figure}[hp]
\centering
\captionsetup{justification=centering,margin=2cm}
\begin{tabular}{cccc}
\\ &
\epsfig{file=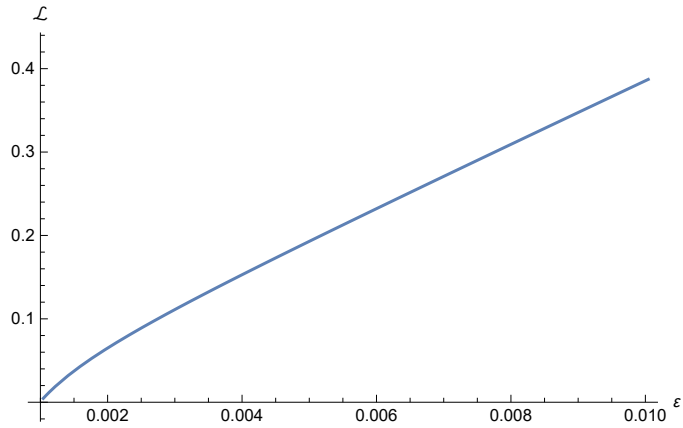,width=0.65\linewidth}\\
\end{tabular}
\caption{ The plot of the proper length $\mathcal{L}$ ($km$) of the gravastar shell against the thickness $\varepsilon$ ($km$) of the shell.\label{fig:THK}}\center
\end{figure}
An increasing proportionate relation between the thickness and length of the shell of the gravastar can be noted in Fig.\ref{fig:THK}.

\subsection{Shell Entropy}

\begin{figure}[hp]
\centering
\captionsetup{justification=centering,margin=2cm}
\begin{tabular}{cccc}
\\ &
\epsfig{file=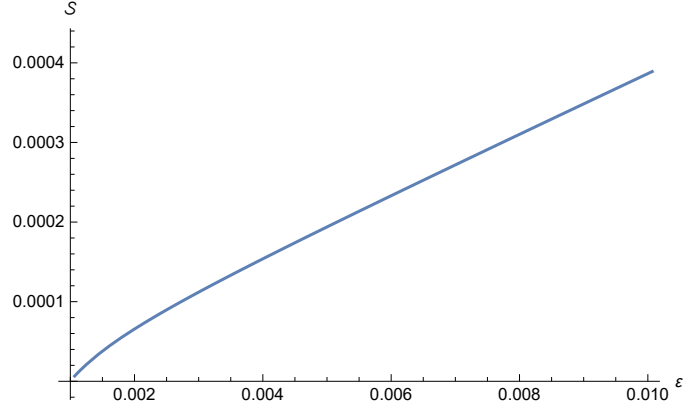,width=0.65\linewidth}\\
\end{tabular}
\caption{ The plot of the within the shell entropy $\mathcal{S}$  against the thickness $\varepsilon$ ($km$) of the shell.\label{fig:V3}}\center
\end{figure}

Mazur and Mottola \cite{MZR} have worked out the zero entropy density in the interior $\mathcal{R}_{1}$ region of the gravastar, which is consistent for the single condensate  region. However, the entropy within the shell can be determined as \cite{GRVFRT}
\begin{equation}\label{Entropy}
  \mathcal{S}=4\pi\int_{\mathcal{D}}^{\mathcal{D}+\varepsilon}r^{2}\mathcal{Z}(r)\sqrt{e^{\varphi}}dr,
\end{equation}
where $\mathcal{Z}(r)$ stands for entropy density defined for the local temperature $\theta(r)$, and is defined as
\begin{equation}\label{LT}
  \mathcal{Z}(r)=\frac{\gamma^{2}K^{2}_{\beta}\theta(r)}{4\pi{h^{2}}}=\gamma\bigg(\frac{K_{\beta}}{h}\bigg)\frac{p}{2\pi},
\end{equation}
where $\gamma$ is a non-dimensional constant. It is to mention here that for current work, we take $c=G=1$ as our geometrical units. Also, if we assume $h=K_{\beta}=1$ as the Planckian units, then the entropy density $\mathcal{Z}(r)$ for the local temperature $\theta(r)$ takes the shape as
\begin{equation}\label{LTP}
  \mathcal{Z}(r)=\gamma\sqrt{\frac{p}{2\pi}}.
\end{equation}
Therefore, under these assumptions, Eq.(\ref{Entropy}) reads
\begin{equation}\label{LTPP}
 \mathcal{S}(\varepsilon)=\gamma\int_{\mathcal{D}}^{\mathcal{D}+\varepsilon}\bigg[\frac{8 \pi ^{3/2} r^2 \sqrt{e^ {\frac{1}{32} C_7 \left(\frac{C_5 \left(4 r-\log \left(C_5-2 C_6 r^2\right)\right)}{C_6}-\frac{2 \sqrt{2} C_5^{3/2} \tanh ^{-1}\left(\frac{\sqrt{2} \sqrt{C_6} r}{\sqrt{C_5}}\right)}{C_6^{3/2}}-2 r^2\right)}}}{\sqrt{\frac{C_5-2 C_6 r^2}{r-2 r^2}}}\bigg]dr.
\end{equation}
The above definite integral seems to be quite complicated. However, if we keep $\mathcal{D}$ fixed, then by the fundamental theorem of calculus, the integral Eq.(\ref{LTPP}) can be transformed to an ODE of the form as
\begin{eqnarray}\label{FTC}
\frac{d\mathcal{S}}{d\varepsilon}=\frac{8\sqrt{e^{\frac{1}{32}C_{7}(-2(\mathcal{D}+\varepsilon)^2-\frac{2\sqrt2
ArcTanh(\frac{\sqrt{2}(\mathcal{D}+\varepsilon)\sqrt{C_{6}}}{\sqrt{C_{5}}})C_{5}^{\frac{3}{2}}}{C_{6}^{\frac{3}{2}}}
+\frac{(4(\mathcal{D}+\varepsilon)-\log(C_{5}-2(\mathcal{D}+\varepsilon)^{2}C_{6}))C_{5}}{C_{6}})}}\pi^{\frac{3}{2}}
(\mathcal{D}+\varepsilon)^{2}}{\sqrt{\frac{C_{5}-2(\mathcal{D}+\varepsilon)^{2}
C_{6}}{(\mathcal{D}+\varepsilon-2(\mathcal{D}+\varepsilon)^2)}}}.
\end{eqnarray}
This equation is a first order highly non-linear ODE. Therefore, through analytic solution of this equation, the explicit expression for $\mathcal{S}$ may not be possible. However, the numerical solution gives us some information which is depicted in Fig.\ref{fig:V3}. This reads the proportional relation of the entropy within the shell $\mathcal{S}$ with the thickness of the shell $\varepsilon$.

\subsection{Energy within the Shell}

\begin{figure}[hp]
\centering
\captionsetup{justification=centering,margin=2cm}
\begin{tabular}{cccc}
\\ &
\epsfig{file=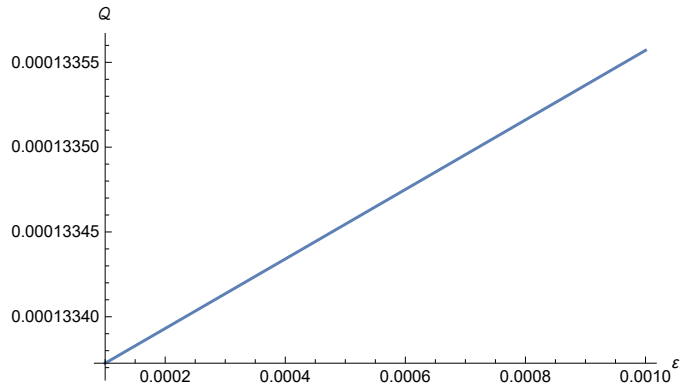,width=0.65\linewidth}\\
\end{tabular}
\caption{ The plot of the energy content $\mathcal{Q}$ ($km$) within the shell against the thickness $\varepsilon$ ($km$) of the shell.\label{fig:V4}}\center
\end{figure}

The energy expression within the shell is given as \cite{GRVFRT}
\begin{equation}\label{EN}
  \mathcal{Q}=4\pi\int_{\mathcal{D}}^{\mathcal{D}+\varepsilon}r^2\rho{dr}=4\pi\int_{\mathcal{D}}^{\mathcal{D}+\varepsilon}r^2\bigg[C_{7}e^{\frac{1}{32}\big(-2r^{2}-\frac{2\sqrt{2}ArcTanh(\frac{\sqrt{2}r\sqrt{C_{6}}}{\sqrt{C_{5}}})C_{5}^{\frac{3}{2}}}{C_{6}^{\frac{3}{2}}}
+\frac{(4r-\log(C_{5}-2r^{2}C_{6}))C_{5}}{C_{6}}\big)}\bigg]dr.
\end{equation}
Applying the similar approach as we do for the integral equation (\ref{LTPP}), we have the transformed differential equation of the form
\begin{eqnarray}
0=\frac{d\mathcal{Q}(\varepsilon)}{d\varepsilon}-4e^{\frac{1}{32}\big(-2(\mathcal{D}+\varepsilon)^2-\frac{2\sqrt2
ArcTanh(\frac{\sqrt{2}(\mathcal{D}+\varepsilon)\sqrt{C_{6}}}{\sqrt{C_{5}}})C_{5}^{\frac{3}{2}}}{C_{6}^{\frac{3}{2}}}
+\frac{(4(\mathcal{D}+\varepsilon)-\log(C_{5}-2(\mathcal{D}+\varepsilon)^{2}C_{6}))C_{5}}{C_{6}}\big)}\pi(\mathcal{D}+\varepsilon)^{2}C_{7}.
\end{eqnarray}
Again, the resulting equation is a highly non-linear differential equation. Applying some appropriate numerical method gives us the information about within the shell energy content with respect to the thickness of the shell. Increasing and proportional positive behavior of the energy-thickness plot is quite apparent from the Fig.\ref{fig:V4}, confirming a non-repulsive approach within the shell. However, within the interior region of the gravastar, we consider the EOS as p = −ρ which shows negative energy region validating the repulsive nature of the interior region.

\section{Concluding Discussion}

The ultimate state of gravitational collapse could lead to the foundation of different matters like neutron stars, black holes, white dwarfs and stripped singularities. This crucial state of the final collapse is a broadly acknowledged field of research in scientific community from different aspects, both theoretical and observational. Though, classical GR suffers from different theoretical problems. One of them is exactly connected to some black hole paradoxical aspects and naked singularities. In order to alleviate such complications, the notion of the existence of compact stars minus event horizons has been recently anticipated, the so-called gravastar: a substitute to black holes. Within these suggested models, a gigantic star in its final phases could conclude its existence as a gravastar which compresses the matter within the gravitational radius $r_{s}=\frac{2GM}{c^{2}}$, that is very close to the Schwarzschild radius but is singularity-free. In this sense, the existence of  quantum vacuum fluctuations are predicted to perform a non-trivial part at or near the event horizon.

In this paper,  we have constructed a stellar model which shares the unique key aspects of the conjecture by Mazur and Mottola and which is also appropriate to be responsive to the whole dynamical analysis. We have introduced a feasible concept of a gravatsar under some particular three-layer structural divisions; the interior of the gravastar is enclosed by a thinly made shell of some stiff ultra-relativistic matter content while the exterior region is entirely vacuum. This is due to the reason that the  Schwarzschild spacetime is perfectly fit to describe the outer region of  the system. As far as the formation of the shell is concerned, it is thought to be extremely thin containing a finite width within restrictions of $r_{1}=\mathcal{D}\leq{r}<\mathcal{D}+\varepsilon=r_{2}$, where $r_{1}$ and $r_{2}$ are the interior and the exterior radii, respectively of the gravastar. These regions, in fact, are structured on the basis of conditions on the EoS parameter as follows: Interior region $\mathcal{R}_1\equiv{0\leq{r}<r_{1}}$, with $\rho=-p$, the shell region $\mathcal{R}_2\equiv{r_{1}}\leq{r}\leq{r_{2}}$, with $+\rho=p$, and the exerior region $\mathcal{R}_3\equiv{r_{2}<r}$, with $\rho=p=0$.
With the consideration of the above mentioned geometrical structure for our investigation under $f(\mathcal{G},T)$ gravity stellar model, the following  prominent features are concluded explicitly as \par
We have shown the correspondence between the density and pressure against the radius $r$ of gravastar with the help of the Fig.\ref{fig:RPr} and it is noted that $\rho\propto{\varepsilon}$. This further points out that the ultra-relativistic fluid is stiff and denser at the outer boundary as compared to the inner boundary of the shell.
We have also worked out an expression for the EoS parameter consisting of factors involving the fractions with square root terms. It is required here to introduce some bounds on these terms to have real expression for $\omega(\mathcal{D})$. By applying some specific conditions such as $\frac{M}{\mathcal{D}}\ll1$ and $C_{1}-\mathcal{D}^{2}>0$, the first order binomial expansion of Eq.$(\ref{EoSp})$ may result into positive and the negative values of the EoS parameter. This may lead us to some particular EoS parameter for the transition layer that could result into the stability of this stellar model.

The graph of proper length $\mathcal{L}$ of the gravastar shell against $\varepsilon$, the thickness of the shell is shown in Fig.\ref{fig:THK}, which depicts its gradually increasing positive behavior. We have also worked out the shell entropy $\mathcal{S}$ after converting the relation into a highly non-linear differential equation. The solution through numerical method has been plotted with respect the thickness $\varepsilon$ of the shell as shown in Fig.\ref{fig:V3}. This plot clearly shows that within the shell entropy $\mathcal{S}$ is increasing gradually with the increasing $\varepsilon$. A directly proportional relation between the energy content $\mathcal{Q}$ and the thickness of the shell $\varepsilon$ has been noted from the plot of $\mathcal{Q}-\varepsilon$ as shown in Fig.$\ref{fig:V4}$.
It is important to mention here that our results under $f(\mathcal{G},T)$ gravity model have been very consistent. It is mentioned here that inclusion of the coupling parameter term $\lambda$ in our $f(\mathcal{G},T)$ gravity model establishes a clear difference and plays a vital role in concluding the key features of this exclusive stellar model. It is concluded here that our $f(\mathcal{G},T)$ stellar model seems to be quite capable of producing very useful and consistent results related to the gravastars. Moreover, it would be interesting to investigate the gravitational lensing effects of gravastars in $f(\mathcal{G},T)$ gravity.
\\\\
\textbf{Acknowledgement}\\\\ Many thanks to the anonymous reviewer
for valuable comments and suggestions to improve the paper.
This work was supported by National University
of Computer and Emerging Sciences (NUCES), Pakistan.


\begin{thebibliography}{40}


\bibitem{MZR}Mazur, P. and Mottola, E.: Proc. Natl. Acad. Sci. U S A. \textbf{101} (2004) 9545; arXiv:gr-qc/0109035.
\bibitem{PRD}Chirenti, C. B. M. H. and  Rezzolla, L.: Phys. Rev. D \textbf{78} (2008) 084011.
\bibitem{PRDM}Wald, R. M.:  Living Rev. Rel. \textbf{4} (2001) 6.
\bibitem{PRD1}Vachaspati, T. : arXiv:0706.1203v2[astro-ph].
\bibitem{PRD2}Chapline, G., Hohlfeld, E., Laughlin, R. B. and Santiago, D. I.: Int. J. Mod. Phys. A \textbf{18} (2003) 3587.
\bibitem{PRD3}Horvat, D., Ilijic, S. and Marunovic, A. Class. Quant. Grav. \textbf{26} (2009) 025003.
\bibitem{PRD4}Copeland, E. J., Sami, M. and Tsujikawa, S.: Int. J. Mod. Phys. D \textbf{15} (2006) 1753.
\bibitem{PRD5}Broderick, A. E. and  Narayan, R.: Class. Quant. Grav. \textbf{24} (2007) 659.


\bibitem{Visser3}Visser, M. and Wiltshire, D. L.: Class. Quant. Grav. \textbf{21} (2004) 1135.
\bibitem{PRD8}Cardoso, V., Pani, P., Cadoni, M. and  Cavaglia, M.: Phys. Rev. D \textbf{77} (2008) 124044.

\bibitem{Carter2}Carter, B. M. N.: Class. Quant. Grav. \textbf{22} (2005) 4551.

\bibitem{Bilic} Bilic´, N., Tupper, G. B. and  Viollier, R. D.: J. Cosmol. Astropart. Phys. \textbf{02} (2006) 013.
\bibitem{LoBO9}Lobo, F. S. N.: Class. Quant. Grav. \textbf{23} (2006) 1525.
\bibitem{LoBO10}Lobo, F. S. N. and Arellano, A. V. B.: Class. Quant. Grav. \textbf{24} (2007) 1069.

\bibitem{PRD6}Kubo, T. and Sakai, N.: Phys. Rev. D \textbf{93} (2016) 084051.
\bibitem{PRD7}Alhamzawi, A.  and Alhamzawi, R.:  Int. J. Mod. Phys. D \textbf{25} (2016) 1650020.
\bibitem{PRD13}Chirenti, C. B. M. H. and Rezzolla, L.: Class. Quant. Grav. \textbf{24} (2007) 4191; Phys. Rev. D \textbf{78}  (2008) 084011.
\bibitem{PRD14}Pani, P., Berti, E., Cardoso, V.,  Chen, Y.  and Norte, R.:  Phys. Rev. D \textbf{80}  (2009) 124047.

\bibitem{Catt}Cattoen, C., Faber, T. and Visser, M.: Class. Quant. Grav. \textbf{22} (2005) 4189.
\bibitem{DeBen}DeBenedictis, A., Horvat, D., Iliji´c, S., Kloster, S. and Viswanathan, K.S.: Class. Quant. Grav.\textbf{23} (2006) 2303.
\bibitem{HVRT}Horvat, D. and Iliji´c, S.: Class. Quant. Grav. \textbf{24} (2007) 5637.
\bibitem{Chirenti2}Chirenti, C. B. M. H. and  Rezzolla, L.: Class. Quant. Grav. \textbf{24} (2007) 4191.
\bibitem{Rocha}Rocha, P., Chan, R.,  da Silva, M. F A. and  Wang, A.: J. Cosmol. Astropart. Phys. \textbf{11} (2008) 010.
\bibitem{Lobo8}Lobo, F. S. N. and  Garattini, R.: J. High Energy Phys. \textbf{12} (2013) 065.
\bibitem{Rahaman3} Rahaman, F., Chakraborty,S. Ray, S., Usmani, A. A., and Islam, S.:Int. J. Theor. Phys. \textbf{54} (2015) 50.

\bibitem{HarkoA}Harko, T.,  Lobo, F. S. N.,  Nojiri, S. and  Odintsov, S. D.:Phys. Rev. D \textbf{84} (2011) 024020.

\bibitem{NJO}Nojiri, S. and Odintsov, S. D.: Phy. Lett. \textbf{599} (2004) 137.
\bibitem{NJO7}Nojiri, S. and Odintsov, S. D.: Phys. Rep. \textbf{505} (2011) 59.

%
\bibitem{NJO8}Nojiri, S. and Odintsov, S. D.: Int. J. Geom. Meth. Mod. Phys. \textbf{04} (2007) 115.
%
%
\bibitem{Capoz}Capozziello, S., Laurentis, M. D. and Odintsov, S. D.: Eur. Phys. J. C \textbf{72} (2012) 2068.

\bibitem{Capoz1}Capozziello, S., Laurentis, M. D. Odintsov, S. D. and Stabile, A.: Phys. Rev. D \textbf{83} (2011) 064004.

\bibitem{Astas}Astashenok, A. V., Capozziello, S., Laurentis, M. D. and Odintsov, S. D.: J. Cosmol. Astropart. Phys. \textbf{1501} (2015) 001.

\bibitem{Capoz2}Capozziello, S., Laurentis, M. D.,  Farinelli, R., and Odintsov, S. D.: Phys. Rev. D \textbf{93} (2016) 023501.
\bibitem{Fel}Felice, A.D. and Tsujikaswa, S.: Living Rev. Rel. \textbf{13} (2010) 3.

\bibitem{Fel2}Bamba, K., Capozziella, S., Nojiri, S. and Odintsov, S. D.: Astrophys. Space Sci. \textbf{342} (2012) 155.

\bibitem{Sta}Starobinsky, A. A.: J. Exp. Theor. Phy. \textbf{86} (2007) 157.
%
%
%
\bibitem{sharif.ayesha} Sharif, M. and Ikram, A.: Eur. Phys. J. C \textbf{76} (2016) 640.




\bibitem{Sir&M}Shamir, M. F. and Ahmad, M.: Eur. Phys. J. C \textbf{77} (2017) 55.

\bibitem{Sir&Me2}Shamir, M. F. and Ahmad, M.: Mod. Phys. Lett. A \textbf{32} (2017) 1750086.
\bibitem{GRVFRT}Das, A., Ghosh, S. Guha, B. K., Das, S., Rahaman, F. and Ray, S.: Phys. Rev. D \textbf{95} (2017) 124011.
\bibitem{Sir&Me3}Shamir, M. F. and Ahmad, M.: Eur. Phys. J. C \textbf{77} (2017) 674.
\bibitem{Sir&SZ}Shamir, M. F. and Zia, S.: Eur. Phys. J. C \textbf{77} (2017) 448.
\bibitem{17}Cognola, G., Elizalde, E., Nojiri, S., Odintsov, S. D. and Zerbini, S.: Phys. Rev. D  \textbf{75} (2007) 086002.
\bibitem{PRD9}Bardeen, J. M.,  Carter, B. and Hawking, S. W.: Commun. Math. Phys. \textbf{31} (1973) 161.
\bibitem{PRD10}Hawking, S. W.: Commun. Math. Phys. \textbf{43} (1975) 199; \textbf{46} (1976) 206.
\bibitem{PRD11}Bekenstein, J. D.:  Phys. Rev. D \textbf{7} (1973) 2333.
\bibitem{PRD12}Jacobson, T.: Phys. Rev. Lett. \textbf{75}  (1995) 1260.
\bibitem{OV}Oppenheimer, J.R. and Volkoff, G.: Phys. Rev. \textbf{55}(1939) 374.
\bibitem{27a}Cooney, A., DeDeo, S., and Psaltis, D.: Phys. Rev. D \textbf{82}(2010) 064033.

\bibitem{31b}Ganguly, A., Gannouji, R., Goswami, R., and Ray, S.: Phys. Rev. D \textbf{89}(2014) 064019.

\bibitem{DM}Momeni, D. and Myrzakulov, R.: Int. J. Geom. Methods Mod. Phys.\textbf{12} (2015) 1550014.

\bibitem{Ast}Astashenok, A. V., Capozziello, S., Laurentis, M. D. and Odintsov, S. D.: J. Cosmol. Astropart. Phys. \textbf{1312} (2013) 040.
\bibitem{PRD21}Moraes, P. H. R. S.,  Arba˜nil, J. D. V. and Malheiro, M.:J. Cosmol. Astropart. Phys. \textbf{06} (2016) 005.
\bibitem{PRD22}Tooper, R. F.: Astrophys. J. \textbf{140} (1964) 434.
\bibitem{PRD23}Ray, S., Espíndola, A. L. M.,  Malheiro, Lemos, J. P. S.  and Zanchin, V. T.: Phys. Rev. D \textbf{68} (2003) 084004.
\bibitem{PRD24}Arbañil, J. D. V., Lemos, J. P. S. and Zanchin, V. T.: Phys. Rev. D \textbf{88} (2013) 084023.
\bibitem{Darmois4}Darmois, G., M´emorial: \textit{des sciences math´ematiques XXV, Fasticule XXV}, Gauthier-Villars, Paris, France, (1927).
\bibitem{Israel2} Israel, W.:Nuovo Cimemto \textbf{44} (1966) 1.
\bibitem{Lanczos}Lanczos, K.: Ann. Phys. (Berlin) \textbf{379} (1924) 518.
\bibitem{67}Rahaman, F., Kalam, M. and Chakraborty, S.:  Gen. Relativ. Gravit. \textbf{38} (2006) 1687.


%

%
%
%
%
%
%
%
%
%
%
%
%
%
%
%
%
%
%
%




\end{thebibliography}
\end{document}